\documentclass[a4paper]{revtex4}
\usepackage[top=1in, bottom=1in, left=1in, right=1in]{geometry}
\usepackage{amsmath}
\usepackage{amssymb}
\usepackage{amsfonts}
\usepackage{graphicx,bm}
\usepackage{dcolumn}
\usepackage{epsfig}

\def \lleq {\lower0.9ex\hbox{ $\buildrel < \over \sim$} ~}
\def \ggeq {\lower0.9ex\hbox{ $\buildrel > \over \sim$} ~}

\def \beq  {\begin{equation}}
\def \eeq  {\end{equation}}
\def \ber  {\begin{eqnarray}}
\def \eer  {\end{eqnarray}}

\newcommand{\be}{\begin{equation}}
\newcommand{\ee}{\end{equation}}
\newcommand{\ba}{\begin{eqnarray}}
\newcommand{\ea}{\end{eqnarray}}
\newcommand{\bea}{\begin{eqnarray*}}
\newcommand{\eea}{\end{eqnarray*}}

\begin{document}

\title{Galileons, phantom fields, and the fate of the Universe}
\author{M. Shahalam$^1$\thanks{%
E-mail address: mohdshahamu@gmail.com}, S. K. J. Pacif$^2$\thanks{%
E-mail address: shibesh.math@gmail.com}, R. Myrzakulov$^3$\thanks{%
E-mail address: rmyrzakulov@gmail.com}}
\affiliation{$^1$Department of Physics, Aligarh Muslim University, Aligarh, India\\
$^2$Center For Theoretical Physics, Jamia Millia Islamia, New Delhi, India\\
$^3$Department of General and Theoretical Physics, Eurasian National
University, Astana, Kazakhstan}

\begin{abstract}
In this paper we study cosmological dynamics of phantom as well as
non-phantom fields with linear potential in presence of Galileon correction $%
(\partial_\mu\phi \partial^\mu\phi) \Box \phi$. We show that the Big Crunch
singularity is delayed compared to the standard case; the delay crucially
depends upon the strength of Galileon correction. As for the phantom
Galileon, $\rho_{\phi}$ is shown to grow more slowly compared to the
standard phantom delaying the approach to singularity. In case, $V\sim
\phi^n, n>4$, Big Rip is also delayed, similar phenomenon is shown to take
place for potentials steeper than the exponential.
\end{abstract}

\maketitle

\section{Introduction}

\label{intro} Observations in cosmology have recently led to confirmation
that the Universe is undergoing an accelerated phase of expansion at present 
\cite{observation1, observation2}. The direct support for the phenomenon
came from the observations of supernovae of type Ia (SNe Ia) \cite%
{observation1}. The explosions of these SNe Ia look fainter than expected in
the Einstein de-Sitter model. The concept of ``\textit{dark energy}'' was
introduced to explain the luminosity-redshift observations of these type Ia
supernovae by modifying the right hand side of Einstein field equations
which give rise to an accelerated expansion of the Universe and thus
explains the unexpected faintness of the supernovae. The weird form of
energy yet remains to be mysterious as there is no direct observational test
to probe it but this is generally assumed that it has a large negative
pressure \cite{ms06}.

In past few years there have been a number of activities for modelling dark
energy including the models with the scalar field and brane world etc. To
this effect, a large varieties of scalar field models are discussed in the
literature including quintessence \cite{quintess}, K-essence \cite{kessen},
spintessence \cite{spint}, tachyon \cite{tachy}, quintom \cite{quintom,od2},
chameleon \cite{chameleon} and many more. These models of scalar field give
the equation of state parameter $w \geq -1$. It is interesting to note that
the observational data also allows models of dark energy with equation of
state parameter crossing $-1$ line (called phantom field models). Thus, a
number of phantom models have been discussed in the literature\cite%
{phantom,ps,gw,od1,at}, for instance, brane world and non-minimally coupled
scalar field models can give phantom energy \cite{vs, msprd, msmnras}. The
simplest way to introduce the phantom effect is provided by a scalar field
with negative kinetic energy term which could be motivated from S-brane
constructs in string theory \cite{stt}. The concept of phantom field was
first used in steady state theory of Hoyle and subsequently incorporated in
Hoyle and Narlikar theory of gravitation \cite{nr}.

The future singularity termed as ``Big Rip'' \cite{aa00} naturally arises in
models with $w<-1$ and is characterized by the divergence of the scale
factor after a finite interval of time. It is generic to keep $w$ as time
dependent rather than to consider it as a constant. This choice of $w$
generates specific scalar field models to avoid the cosmic doomsday \cite%
{ps,Carroll} which requires a particular class of phantom field potentials.

There are alternative ways to explain the accelerated expansion by modifying
the left hand side of Einstein field equations \textit{a la} modified
theories of gravity. Following this, a special class of dark energy model
based on the large scale modification of gravity called \textit{Galileon
gravity }\cite{Galileon,alig10} was proposed. The distinguished feature of
this theory is that it provides a consistent modification of general
relativity leaving the local physics intact. This modified gravity in this
scheme can give rise to the observed late time cosmic acceleration and also
it is free from negative energy instabilities. The Galileon field has five
field Lagrangians $\mathcal{L}_{i}$ ($i=1,...,5$) in 4-dimensional
space-time. The Lagrangian $\mathcal{L}_{1}$ is linear, $\mathcal{L}_{2}$ is
the standard kinetic term and $\mathcal{L}_{3}$ represents the Vainshtein
term consisting of three Galileon fields that is related to the decoupling
limit of Dvali, Gabadadze, and Porrati (DGP) model \cite{DGP} while $%
\mathcal{L}_{4}$ and $\mathcal{L}_{5}$ consists of higher order non-linear
derivative terms of field. In case, we study scalar field with linear
potential, it becomes obligatory to compliment it by the higher derivative
Galileon terms in the Lagrangian. For simplicity, we shall consider the
lowest Galileon term $\mathcal{L}_{3}$ for phantom and non-phantom fields
with linear potential. On purely phenomenological grounds, we also examine
the phantom case with a general potential term $V(\phi)$ \cite%
{alig12,walig12,msgrg} complimented by Galileon term. In this case, we focus
on some general features of cosmological dynamics, in particular, current
acceleration and future evolution of the Universe.

Recently, it was found that in quintessence models where scalar field
potentials turn negative might lead to collapse of the Universe in distant
future \cite{c1,c2,c3,c4,c5} dubbed Big Crunch singularity. Lykkas and
Perivolaropoulos show that the cosmic doomsday singularity can be avoided in
Scalar-Tensor Quintessence with a linear potential by some values of the
non-minimal coupling parameter \cite{peri}. In this paper, we shall examine
these and other aforesaid issues in presence of Galileon correction $%
\mathcal{L}_3$ in the Lagrangian.

The paper is organized as follows. In Section II, we consider Galileon field
model with linear potential which is generically non-minimally coupled
scalar field model and investigates present and future evolution of the
Universe for both phantom and non-phantom cases. In Section III, we consider
Galileon phantom field model with steep exponential potential and examine
the future evolution of the Universe. We summarize our results in the
Section IV.

\section{Galileon field with linear potential}

In this section, we consider Galileon field action possessing up to the
third order term in the Lagrangian with a field potential $V(\phi )$ \cite%
{alig12,walig12,msgrg}. 
\begin{equation}
S=\int d^{4}x\sqrt{-g}\left[ \frac{M_{pl}^{2}}{2}R-\frac{1}{2}\epsilon
\left( \triangledown \phi \right) ^{2}-\frac{\beta }{2M^{3}}\left(
\triangledown \phi \right) ^{2}\square \phi -V\left( \phi \right) \right]
+S_{m}  \label{eq:action}
\end{equation}%
where, $\epsilon $ = -1 and +1, for phantom and non-phantom Galileon fields
respectively. $M_{pl}^{2}=1/{8\pi G}$ is the reduced Plank mass and the
constant $\beta $ is dimensionless. $S_{m}$ entitles the matter action and $%
M $ is a constant of mass dimension one. For simplicity, we fix here $%
M=M_{pl}$. In a homogeneous isotropic flat
Friedmann-Lemaitre-Robertson-Walker (FLRW) Universe, the equations of motion
are obtained by varying the action (\ref{eq:action}) with respect to metric
tensor $g_{\mu \nu }$ and scalar field $\phi $ as

\begin{equation}
3M_{pl}^{2}H^{2}=\rho _{m}+\frac{1}{2}\ \epsilon \dot{\phi}^{2}-3\frac{\beta 
}{M_{\mathrm{pl}}^{3}}H{\dot{\phi}}^{3}+V{(\phi )}\,,  \label{eq:Hd}
\end{equation}%
\begin{equation}
M_{pl}^{2}(2\dot{H}+3H^{2})=-\frac{1}{2}\ \epsilon \dot{\phi}^{2}-\frac{%
\beta }{M_{\mathrm{pl}}^{3}}\dot{\phi}^{2}\ddot{\phi}+V(\phi ),  \label{eq:H}
\end{equation}%
\begin{equation}
\frac{\ddot{a}}{a}=-\frac{1}{6M_{pl}^{2}}\Bigl(\rho _{m}+2\epsilon \dot{\phi}%
^{2}-\frac{3\beta }{M_{pl}^{3}}(H\dot{\phi}^{3}-\dot{\phi}^{2}\ddot{\phi}%
)-2V(\phi )\Bigr),  \label{eq:add1}
\end{equation}%
\begin{equation}
\epsilon \ddot{\phi}+3H\epsilon \dot{\phi}-3\frac{\beta }{M_{\mathrm{pl}}^{3}%
}\dot{\phi}\Bigl(3H^{2}\dot{\phi}+\dot{H}\dot{\phi}+2H\ddot{\phi}\Bigr)%
+V^{\prime }(\phi )=0,  \label{eq:phidd}
\end{equation}%
where, $\prime $ denotes derivative with respect to $\phi $ and $\rho _{m}$
is the energy density of matter. The energy conservation equation of matter
is given by 
\begin{equation}
\dot{\rho}_{m}+3H\rho _{m}=0.  \label{eq:rhom}
\end{equation}%
%
%
%
%
%
%
%
%
%
%
%
%
%
%
%
%
%
In the radiation/matter dominated phase, the Universe is dominated by a
perfect fluid with equation of state $p=w\rho $. In this phase of evolution
the density of matter $\rho _{m}$ dominates over the field $\phi $. 
With the expansion of the Universe over time the Hubble parameter $H$ begins
decreasing and the scalar field $\phi $ starts dominating. The total energy
content of the Universe $\rho _{total}=\rho _{m}+\rho _{\phi }\simeq \rho
_{\phi }=\frac{1}{2}\epsilon \dot{\phi}^{2}+V(\phi )-\frac{3\beta }{%
M_{pl}^{3}}H\dot{\phi}^{3}$. Therefore equation (\ref{eq:Hd}) reduces to 
\begin{equation}
3M_{pl}^{2}H^{2}=\frac{1}{2}\ \epsilon \dot{\phi}^{2}+V(\phi )-\frac{3\beta 
}{M_{pl}^{3}}H\dot{\phi}^{3},  \label{a}
\end{equation}%
which is difficult to solve analytically. In what follows, we shall solve
the evolution equations numerically and plot the future evolution
graphically.

Introducing the dimensionless parameters 
\begin{eqnarray}
H_{0}t &\longrightarrow &t,  \notag \\
\frac{\phi }{\sqrt{3}M_{pl}} &\longrightarrow &\phi ,  \label{eq:dimension}
\\
\frac{V_{0}}{\sqrt{3}M_{pl}^{2}H_{0}^{2}} &\longrightarrow &V_{0}.  \notag
\end{eqnarray}%
The system of equations (\ref{eq:add1}) and (\ref{eq:phidd}) can be written
as 
\begin{equation}
\frac{\ddot{a}}{a}=-\epsilon \dot{\phi}^{2}+\frac{3\sqrt{3}~\beta }{2}\left( 
\frac{H_{0}}{M_{pl}}\right) ^{2}\left( \frac{\dot{a}}{a}\dot{\phi}^{3}-\dot{%
\phi}^{2}\ddot{\phi}\right) +V_{0}\phi -\frac{\Omega _{m0}}{2a^{3}},
\label{eq:add}
\end{equation}%
\begin{equation}
\epsilon \ddot{\phi}+3\frac{\dot{a}}{a}~\epsilon \dot{\phi}-3\sqrt{3}\beta 
\dot{\phi}\left( \frac{H_{0}}{M_{pl}}\right) ^{2}\left\{ 3\frac{\dot{a}^{2}}{%
a^{2}}\dot{\phi}+\left( \frac{a\text{ }\ddot{a}-\dot{a}^{2}}{a^{2}}\right) 
\dot{\phi}+2\frac{\dot{a}}{a}\ddot{\phi}\right\} +V_{0}=0.  \label{eq:phiddn}
\end{equation}%
The equation of state parameter $w$ for Galileon field is defined as 
\begin{eqnarray}
w &=&\frac{p_{\phi }}{\rho _{\phi }};  \label{eq:eos} \\
p_{\phi } &=&\frac{1}{2}~\epsilon \dot{\phi}^{2}+\frac{\beta }{M_{\mathrm{pl}%
}^{3}}\dot{\phi}^{2}\ddot{\phi}-V(\phi ),  \notag \\
\rho _{\phi } &=&\frac{1}{2}~\epsilon \dot{\phi}^{2}-3\frac{\beta }{M_{%
\mathrm{pl}}^{3}}H\dot{\phi}^{3}+V{(\phi )}.  \notag
\end{eqnarray}%
\begin{figure}[tbp]
\begin{center}
\begin{tabular}{cc}
{\includegraphics[width=2.3in,height=2.1in,angle=0]{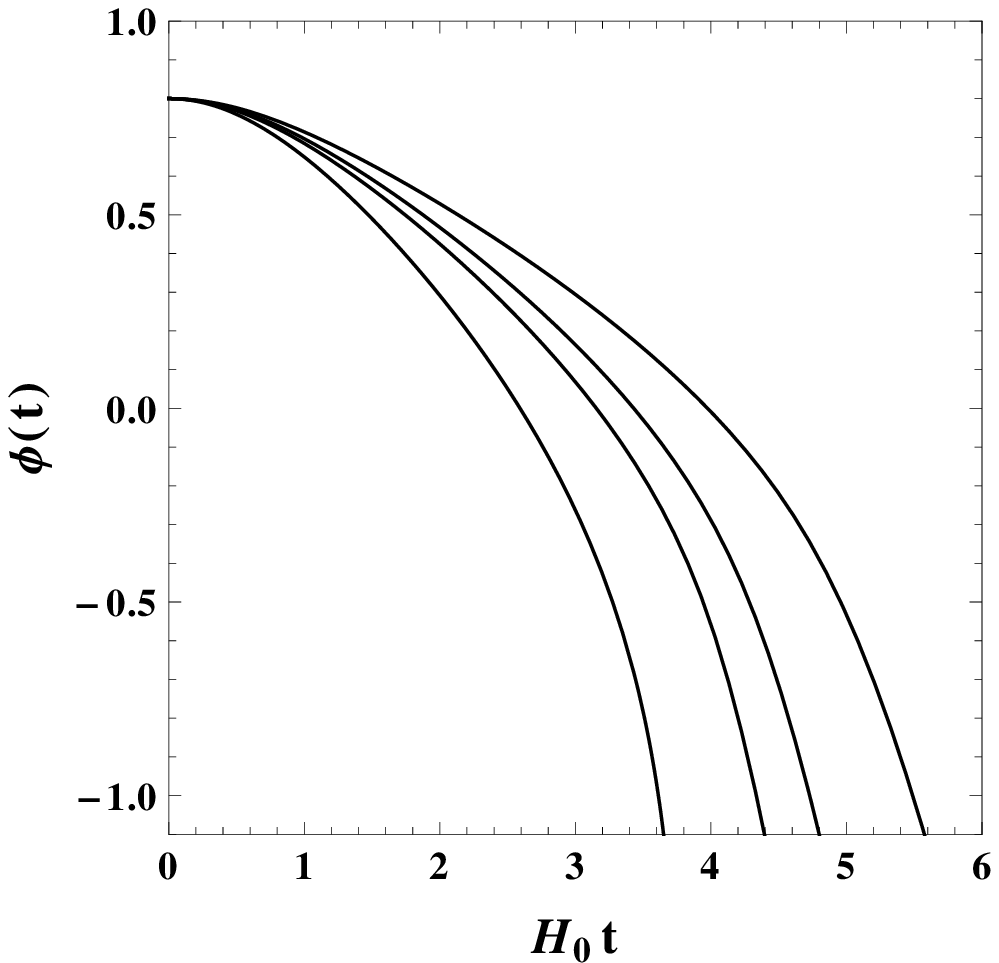}} & {%
\includegraphics[width=2.3in,height=2.1in,angle=0]{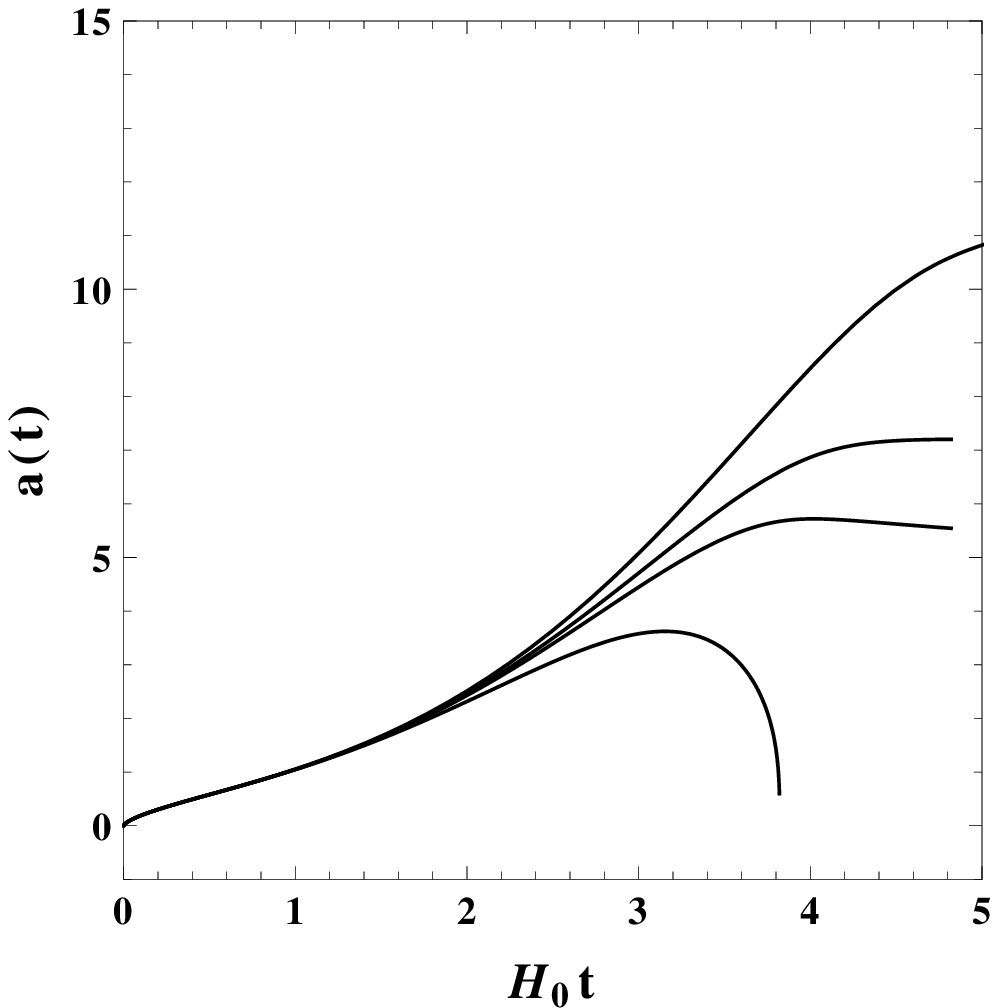}} \\ 
{\includegraphics[width=2.3in,height=2.1in,angle=0]{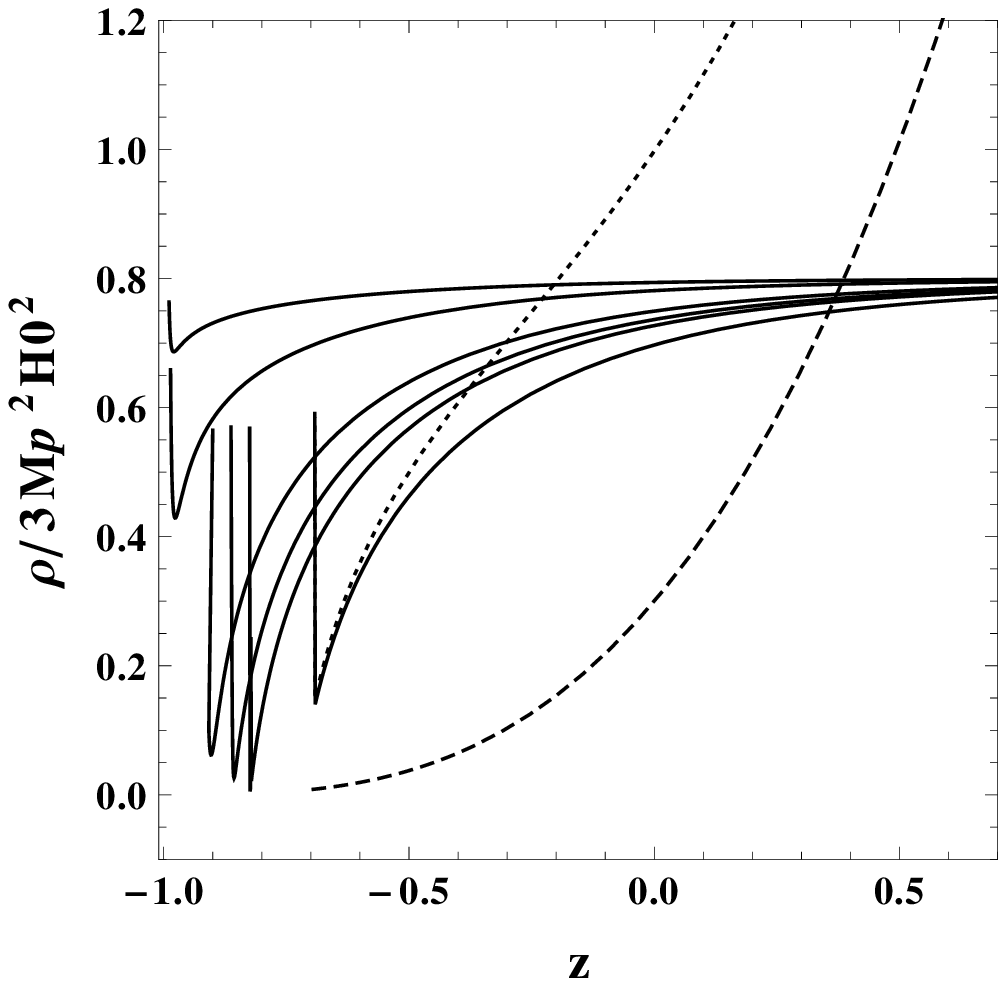}} & {%
\includegraphics[width=2.3in,height=2.1in,angle=0]{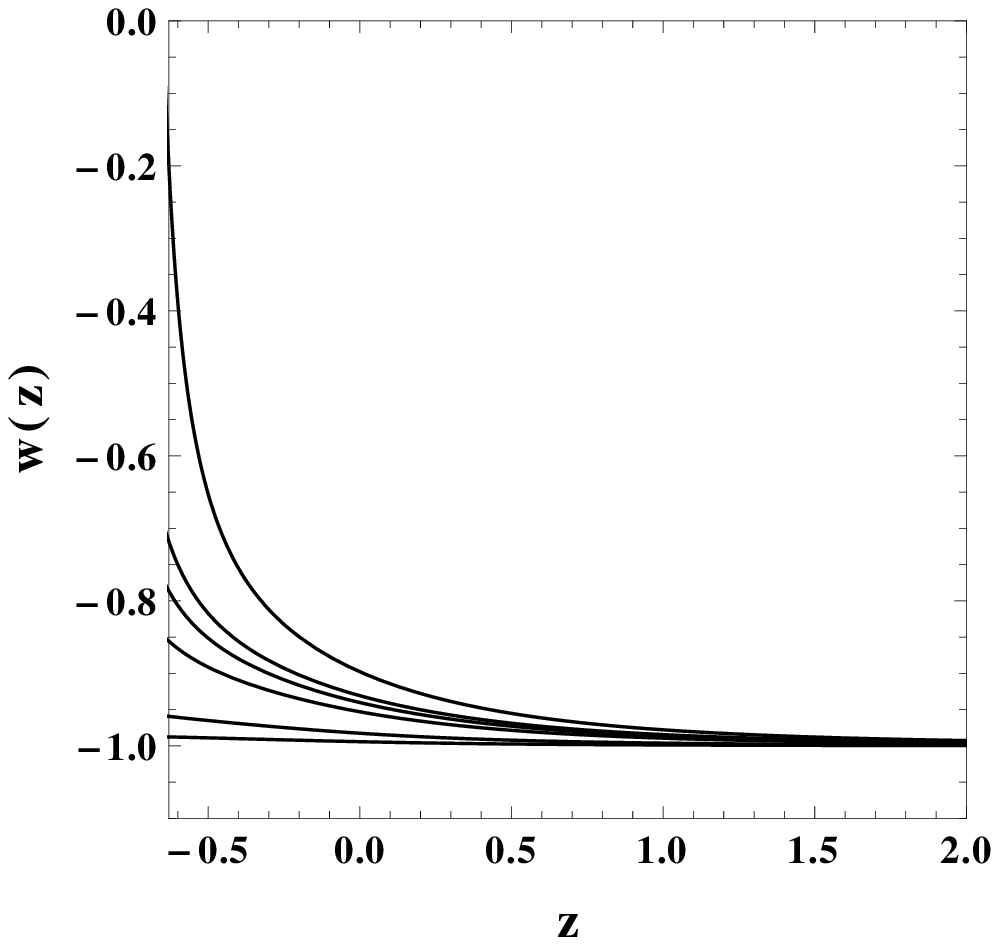}}%
\end{tabular}%
\end{center}
\caption{{\protect\small The upper panels show the evolution of field $%
\protect\phi (t)$ and scale factor $a(t)$ versus time ($H_{0}t$) for
Galileon field having linear potential for different values of $\protect%
\beta $ and $V_{0}=1$ showing collapse nature in future (see upper right
panel). The time is normalized by $H_{0}$ (Hubble constant at present
epoch). The present time corresponds to $t_{0}=0.96$. The left bottom panel
represents the evolution of energy density $\protect\rho $ versus redshift $%
z $. The solid lines correspond to Galileon field for different values of $%
\protect\beta $. The dashed and dotted lines represent the energy density of
matter and total energy density of the Universe respectively with $\protect%
\beta =0$ (quintessence). The right bottom panel shows the evolution of
equation of state parameter $w$ versus redshift $z$ for Galileon field with
different values of $\protect\beta $. In this figure, upper panels are
plotted for $\protect\beta =0,0.3,0.5,1$ (for more higher values, collapse
shifted in far distant future) whereas the lower panels have $\protect\beta %
=0,0.3,0.5,1,10,100$ from bottom to top but in the bottom right panel from
top to bottom.}}
\label{figquint}
\end{figure}
In case of Galileon field model, we are considering two cases phantom and
non-phantom. First we shall discuss non-phantom case.\newline
\newline
\textbf{Case I: Non-phantom ($\epsilon =+1$)} \newline
\newline
When $\beta =0$ the Galileon field action (\ref{eq:action}) reduces to the
standard quintessence field. In equations (\ref{eq:add}) and (\ref{eq:phiddn}%
), we have two variables, namely, scale factor $a$ and field $\phi $. The
term $3\beta H\dot{\phi}^{3}/M_{pl}^{3}$ in equation (\ref{eq:Hd}) is the
Galileon correction term which depends upon $a$, $\phi $ and parameter $%
\beta $. The different values of $\beta $ puts the strength of Galileon
correction term over the quintessence term. For $\beta =0$, the evolution of
Galileon field model is same as the standard quintessence throughout the
history of the Universe. Hence all non zero values of $\beta $ find the
departure from quintessence and also the effect of Galileon correction term.
Therefore in this analysis we take $\beta $ as a model parameter.

Now we solve the equations (\ref{eq:add}) and (\ref{eq:phiddn}) numerically
with the assumption that the field $\phi$ was frozen initially (i.e. $%
\phi(t_i)=\phi_i$ and $\dot{\phi}(t_i)=0$) caused by huge Hubble damping.
This is identical to thawing type of models \cite{scherrer}. We use
following initial conditions ($t\rightarrow t_{i}\simeq 0$) 
\begin{eqnarray}
a(t_{i}) &=&\left( \frac{9\Omega _{0m}}{4}\right) ^{1/3}t_{i}^{2/3}  \notag
\\
\dot{a}(t_{i}) &=& \frac{2}{3}\left( \frac{9\Omega _{0m}}{4}\right)
^{1/3}t_{i}^{-1/3}  \notag \\
\phi (t_{i}) &=&\phi _{i}  \notag \\
\dot{\phi}(t_{i}) &=&0.  \label{eq:init}
\end{eqnarray}
With the above initial conditions and by tuning $\phi_i$, we get the
following parameters at the present time, 
\begin{eqnarray}
a(t_{0}) &=& 1,  \notag \\
H(t_{0}) &=& 1,  \notag \\
\Omega_{0m} &=& 0.3,  \label{eq:parm}
\end{eqnarray}
where, $t_{0}$ is defined as the time when the scale factor is unity. In the
upper panels of figure \ref{figquint}, we present the dynamical evolution of
field $\phi$ and scale factor $a$ for different values of $\beta$ and $V_0=1$%
. For $\beta=0$, the evolution of $a$ is alike to the standard quintessence
model. Initially, the field is positive and the Universe gets expansion with
the late time cosmic acceleration as soon as field changes sign, in future,
potential becomes negative and the scale factor collapses to a Big Crunch
singularity. However, for larger values of $\beta$ the sign of field changes
in more distant future and correspondingly $V(\phi)$ becomes negative.
Therefore collapse of scale factor is shifted in more distant future for
higher values of $\beta$. In other words, the cosmic doomsday is delayed for 
$\beta > 0$.

In the lower left panel of figure \ref{figquint}, we show the evolution of
energy density for various values of $\beta$ and $V_0=1$. Initially, the
Galileon field imitates the $\Lambda$CDM like behaviour and its energy
density is highly sub-dominant to the matter energy density $\rho_m$ and
persists so, for most of the time of expansion. The Galileon field remains
in the state with $w=-1$ till the epoch $\rho_{\phi}$ goes near to $\rho_{m}$%
. At late times, the energy density of Galileon field gets to the matter,
overtakes it and begins decreasing ($w > - 1$), and acquires the present
accelerated expansion of the Universe having $\Omega_{0m} \simeq 0.3$ and $%
\Omega_{0\phi} \simeq 0.7$. Thereafter $\rho_{\phi}$ continuously decreases
until it comes a point where $\phi$ is negative (i.e. $\phi < 0$) and $\dot{%
\phi}^{2}/{2}+V(\phi )-3 \beta H\dot{\phi}^{3}/{M_{pl}^{3}}=0$. Therefore, $%
H\longrightarrow 0$, i.e. the total energy density of the Universe reaches
to zero and bounce occurs. For $\beta=0$, Galileon field behaves as standard
quintessence and the similar behaviour for quintessence is shown in ref. 
\cite{c5}. As we go for higher values of $\beta$ ($\beta=0.3, 0.5, 1, 10,
100 $) the bounce and collapse shifted in distant future. One can say that
the bounce and collapse delayed for higher values of $\beta$.

The evolution of equation of state for $V_0=1$ and various values of $\beta$
is shown in the lower right panel of figure \ref{figquint}. For $\beta=0$,
the equation of state of Galileon field reduces to the equation of state of
standard quintessence and diverges from the equation of state of $\Lambda$%
CDM model. As the values of $\beta$ are increased, we get more and more
deviation in $w$ from the case of standard quintessence and approaches
towards the $\Lambda$CDM model. The higher values of $\beta$ for Galileon
field with linear potential are in good agreement with the observations as
shown in ref. \cite{walig12} where they have imposed observational
constraints on Galileon correction term which is associated with $\beta$. 
\begin{figure}[tbp]
\begin{center}
\begin{tabular}{cc}
{\includegraphics[width=2.3in,height=2.1in,angle=0]{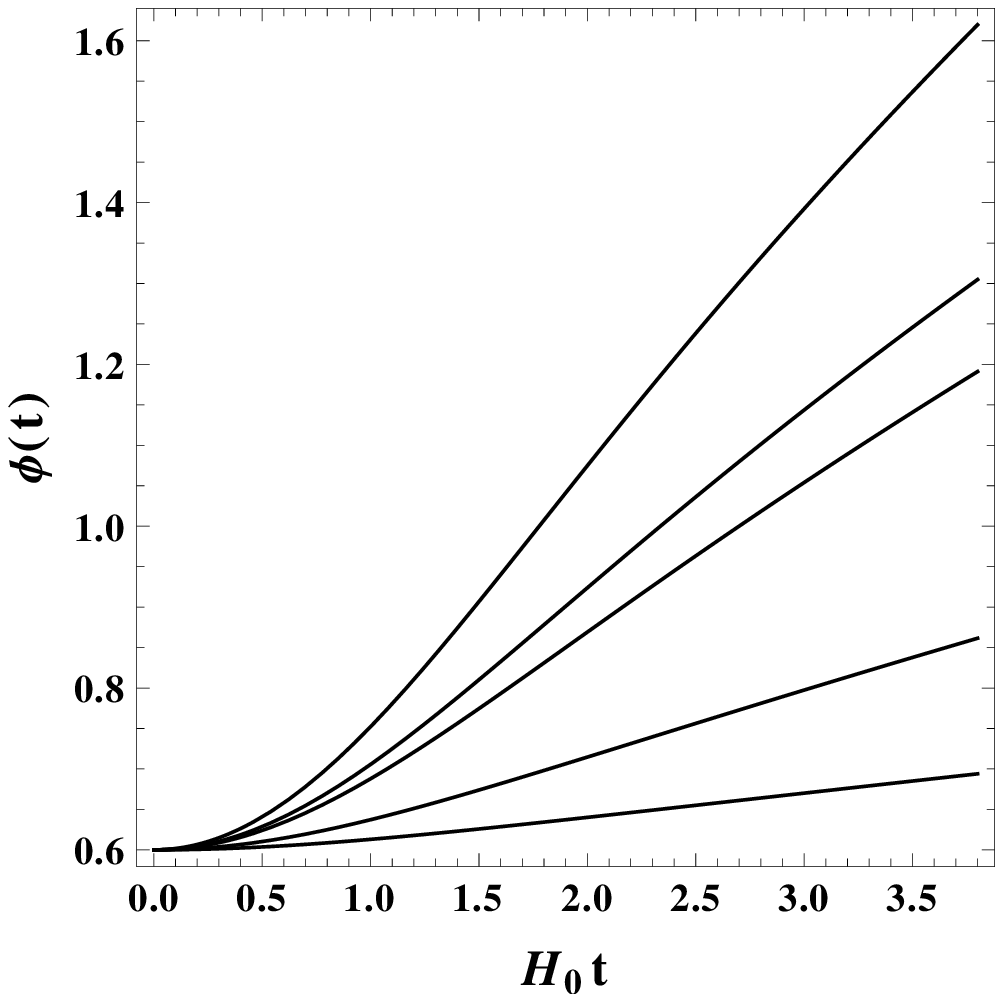}} & {%
\includegraphics[width=2.3in,height=2.1in,angle=0]{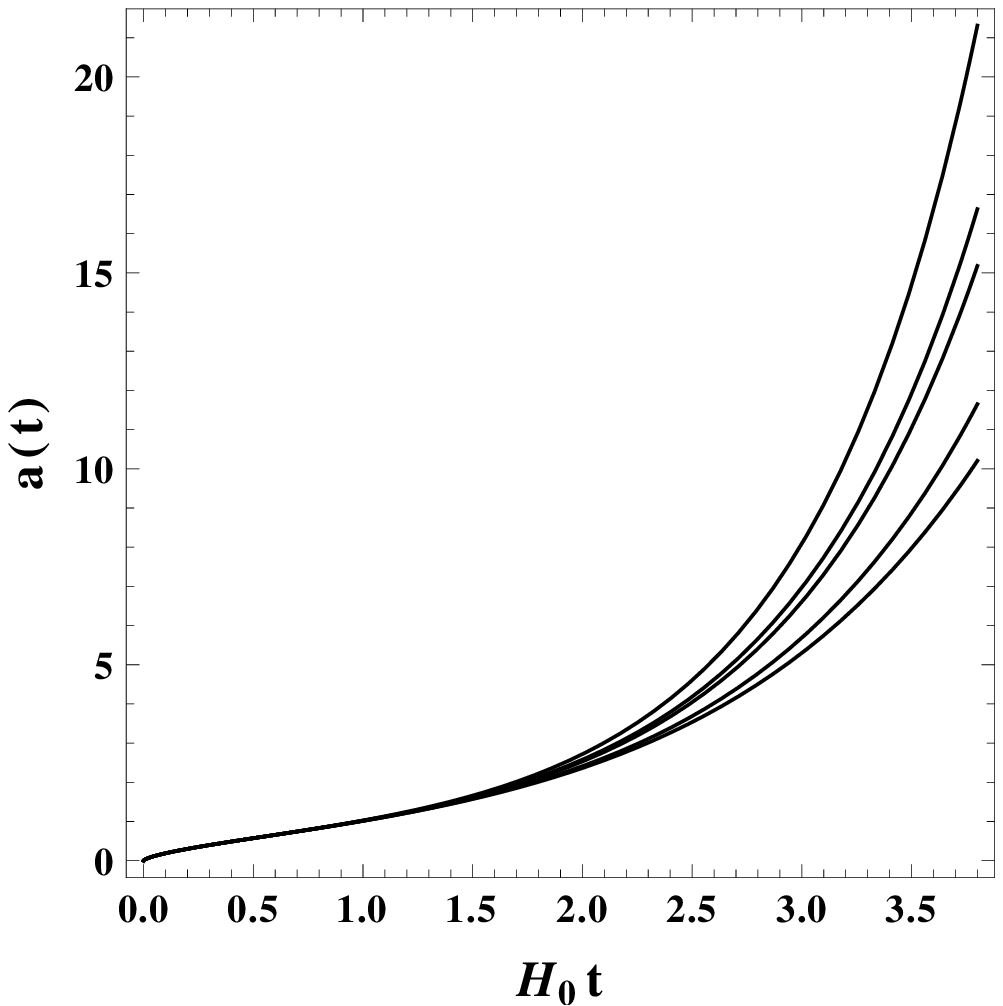}} \\ 
{\includegraphics[width=2.3in,height=2.1in,angle=0]{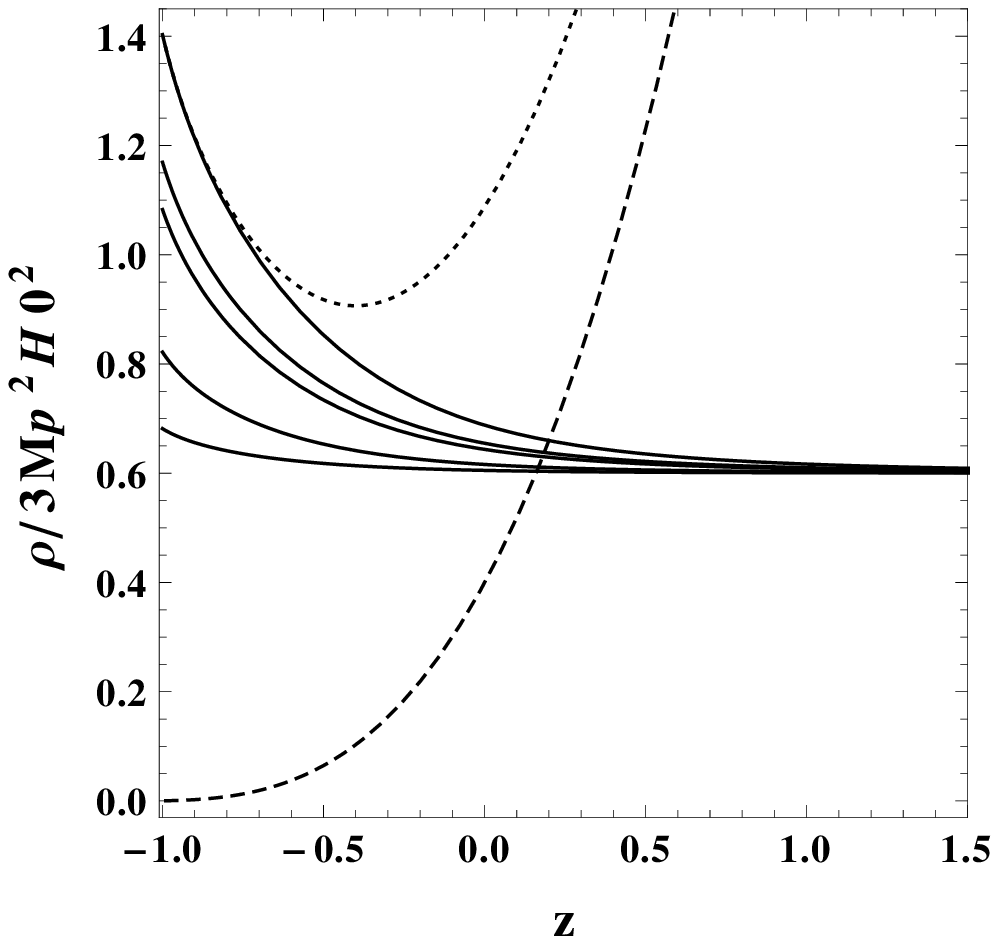}} & {%
\includegraphics[width=2.3in,height=2.1in,angle=0]{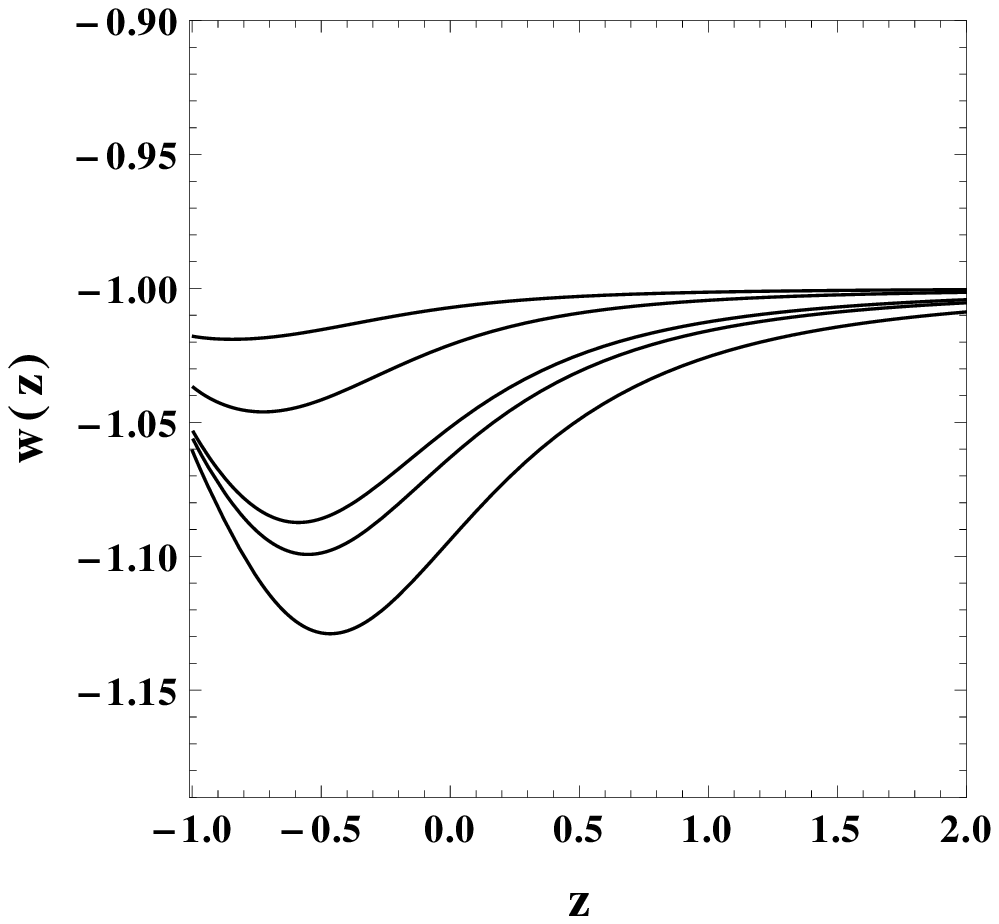} }%
\end{tabular}%
\end{center}
\caption{{\protect\small The evolution of field $\protect\phi(t)$ and scale
factor $a(t)$ versus time ($H_{0}t$) for Galileon phantom field with linear
potential are plotted and shown in the upper panels for different values of $%
\protect\beta$ and $V_0=1$. The time is normalized by $H_{0}$ and the
present time is $t_{0}= 0.96$. The upper right panel shows the divergent
nature of scale factor, in future, after some finite interval of time. The
energy density $\protect\rho$ versus redshift $z$ is shown in the left
bottom panel, where the solid lines correspond to Galileon phantom field for
various values of $\protect\beta$. The dashed and dotted lines represent the
energy density of matter and total energy density of the Universe
respectively for $\protect\beta=0$ (standard phantom). The right bottom
panel shows the evolution of equation of state $w$ versus redshift $z$ for
Galileon phantom field with various values of $\protect\beta$. In this
figure, all the panels have $\protect\beta=0, 0.5, 1, 10$ and 100 from top
to bottom whereas bottom right panel from bottom to top. }}
\label{figphant}
\end{figure}
\newline
\newline
\textbf{Case II: Phantom ($\epsilon=-1$)} \newline
\newline
First we are considering the case of $\beta=0$. Hence, the action (\ref%
{eq:action}) reduces to the action of phantom field minimally coupled to
gravity and matter. 
\begin{equation}
S=\int d^{4}x\sqrt{-g}\left[ \frac{M_{pl}^{2}}{2}R+\frac{1}{2} \left(
\triangledown \phi \right) ^{2} -V\left( \phi \right) \right] +S_{m}
\label{eq:actionp}
\end{equation}%
The wrong sign in the kinetic energy term of equation (\ref{eq:actionp})
gives the ghost field in the context of quantum field theory or phantom
field in cosmology. As a dark energy candidate, the equation of state of
phantom field is marginally favoured by the present observations \cite%
{planck}. The vital cosmological dynamics of phantom field has been broadly
discussed in the literature. However, it is plagued with intense quantum
instabilities. Theoretically, we still do not know the basic origin of $w <
-1$. In the recent past, it has been discussed that the opposite sign in the
kinetic energy term does not give instabilities, required that higher order
derivative terms should be included in the action \cite{nima}.

We take Galileon phantom field model by invoking negative sign in the
kinetic energy term. For $\beta=0$, it behaves as a standard phantom field
model. In this case, the initial kinetic term of the phantom field decreases
due to Hubble damping term in equation (\ref{eq:phiddn}) and as a result the
field freezes for a while till the epoch $\rho_{\phi}$ approaches to $%
\rho_{m}$ (see bottom left panel of figure \ref{figphant}). Eventually, the
field switches on and the future evolution depends upon the shape of the
potential $V(\phi)$.

When $\dot{\phi}$ is nearly frozen and phantom energy starts to dominate,
then the system of equations (\ref{eq:Hd}) and (\ref{eq:phidd}) reduces to
(for $\beta =0$ case) 
\begin{equation}
H^{2}\simeq \frac{V(\phi )}{3M_{pl}^{2}},~~~~~~\dot{\phi}\simeq \frac{%
V^{\prime }(\phi )}{3H}.  \label{b}
\end{equation}%
The ratio kinetic to potential term can be written as 
\begin{equation}
\frac{\dot{\phi}^{2}}{2V(\phi )}=\frac{M_{pl}^{2}}{6}~\frac{V^{\prime 2}}{%
V(\phi )^{2}}=\frac{M_{pl}^{2}}{6}~\frac{1}{\phi ^{2}},  \label{c}
\end{equation}%
where, $V(\phi )=V_{0}~\phi /M_{pl}$, the ratio kinetic energy to potential
energy term is proportional to $1/\phi ^{2}$ and goes to zero as a result
the kinetic energy term remains sub-dominant continually. This is similar to
the slow-roll regime for an ordinary field and can be called as
\textquotedblleft slow climb\textquotedblright\ \cite{at, ref1a}. How to exit from rip was discussed in Ref. \cite{ref1b}. The equation of
state approaches towards $-1$ (see bottom right panel of figure \ref%
{figphant}) with an increasing energy density as shown in bottom left panel
of figure \ref{figphant}. The estimation $\frac{\dot{\phi}^{2}}{2V(\phi )}%
\longrightarrow 0$ is not valid for an exponential and steeper potentials.
We shall discuss this case in Section \ref{sec:exp}. In case of phantom with
Galileon correction (for $\beta \neq 0$), with the domination of phantom
energy and $\dot{\phi}$ is small, the system of equations (\ref{eq:Hd}) and (%
\ref{eq:phidd}) reduces to (by taking the subleading terms) 
\begin{equation}
H^{2}\simeq \frac{V(\phi )}{3M_{pl}^{2}}\text{ },\text{ \ \ \ }\dot{\phi}%
\simeq \frac{M_{pl}^{3}}{6\beta H}\left[ -1\pm \sqrt{1+\frac{4\beta }{%
M_{pl}^{3}}V^{\prime }(\phi )}\right] .  \label{d}
\end{equation}%
In slow roll approximation, the term $\frac{4\beta }{M_{pl}^{3}}V^{\prime
}(\phi )$ is small and we have from (\ref{d}) 
\begin{equation}
\frac{\dot{\phi}^{2}}{2V}\approx \frac{M_{pl}^{2}}{6}\left( \frac{V^{\prime
}(\phi )}{V(\phi )}\right) ^{2}\left[ 1-\frac{2\beta }{M_{pl}^{3}}V^{\prime
}(\phi )\right] ,  \label{e}
\end{equation}%
showing that the\ presence of Galileon correction term enhances the slow
climb for monotonically increasing $V(\phi )$ . Keeping in mind the, $%
\epsilon =-\frac{\dot{H}}{H^{2}}=\frac{3}{2}(1+\omega _{eff}(\phi )),$ we
have shown numerically that the Galileon correction term for large values of 
$\beta $ moves $\omega _{eff}(\phi )$ towards the de-Sitter point, though $%
\omega _{eff}(\phi )$ yet remains to be less than -1.

In the upper panels of figure \ref{figphant}, we show the evolution of
phantom field $\phi $ and scale factor $a$ versus time. The phantom field
and scale factor both diverges after finite interval of time (in future) and
correspondingly energy density of phantom field $\rho _{\phi }$ increases
slowly (see bottom left panel of figure \ref{figphant}). In this case, the
equation of state first decreases from $-1$ and then eventually increases
towards $-1$ and comes near to it asymptotically \cite{at} that is shown in
bottom right panel of figure \ref{figphant}. To this effect, final Universe
would be different from both the de-Sitter and Big Rip, and an infinite time
would be taken to reach an infinite energy density.

When we add Galileon correction term and go for higher values of $\beta$
(0.5, 1, 10, 100), the scale factor shows less divergence nature than the
case of $\beta=0$ and correspondingly $\rho_\phi$ grows slowly as shown in
the upper right and left bottom panels of figure \ref{figphant}; Initially,
the Galileon phantom field imitates the $\Lambda$CDM like behaviour and its
energy density is highly sub-dominant to the matter energy density $\rho_m$
and remains so, for most of the time of evolution. The Galileon phantom
field remains in the state with $w=-1$ till the epoch $\rho_{\phi}$ goes
near to $\rho_{m}$. At late times, the energy density of Galileon phantom
field approaches to matter, overtakes it and begins increasing ($w < - 1$),
and acquires the present accelerated expansion of the Universe having $%
\Omega_{0m} \simeq 0.3$ and $\Omega_{0\phi} \simeq 0.7$. For higher values
of $\beta$ ($\beta= 0.5, 1, 10, 100$) the slow growing divergence shifted
towards lower values of $\rho_\phi$. In the right bottom panel of figure \ref%
{figphant}, we present the evolution of $w$ versus redshift $z$ for Galileon
phantom field. For $\beta=0$, the equation of state of Galileon phantom
field deviates more from the equation of state of $\Lambda$CDM model. As the
values of $\beta$ are increased, we get less deviation from $\Lambda$CDM.
For all values of $\beta$, the equation of state first decreases from $-1$
and then subsequently increases towards $-1$ and comes near to it
asymptotically. Hence, we get smaller deviation in equation of state
parameter $w$ from $\Lambda$CDM for $\beta > 0$. The effect of the Galileon correction on the evolution of phantom field was also studied in Ref. \citep{ref1c}.
\begin{figure}[tbp]
\label{figexp}
\begin{center}
\begin{tabular}{cc}
{\includegraphics[width=2.3in,height=2.1in,angle=0]{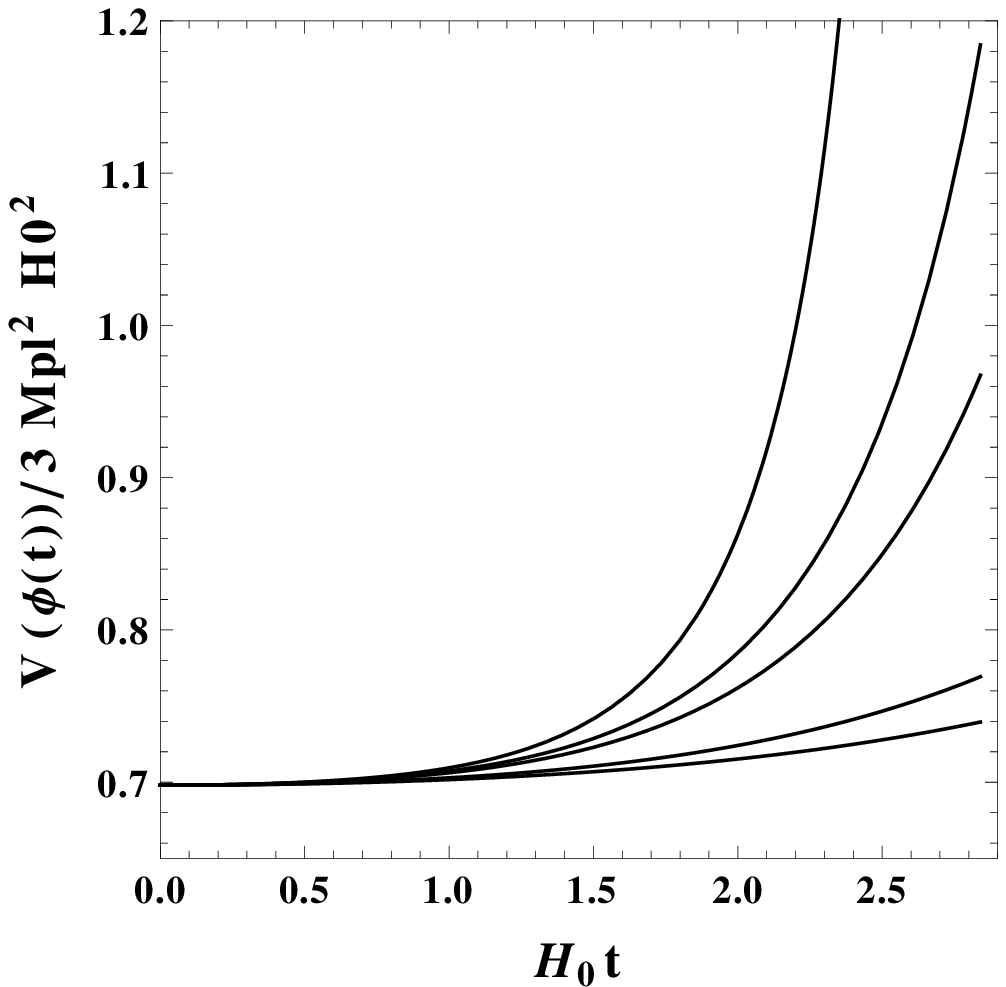}} & {%
\includegraphics[width=2.3in,height=2.1in,angle=0]{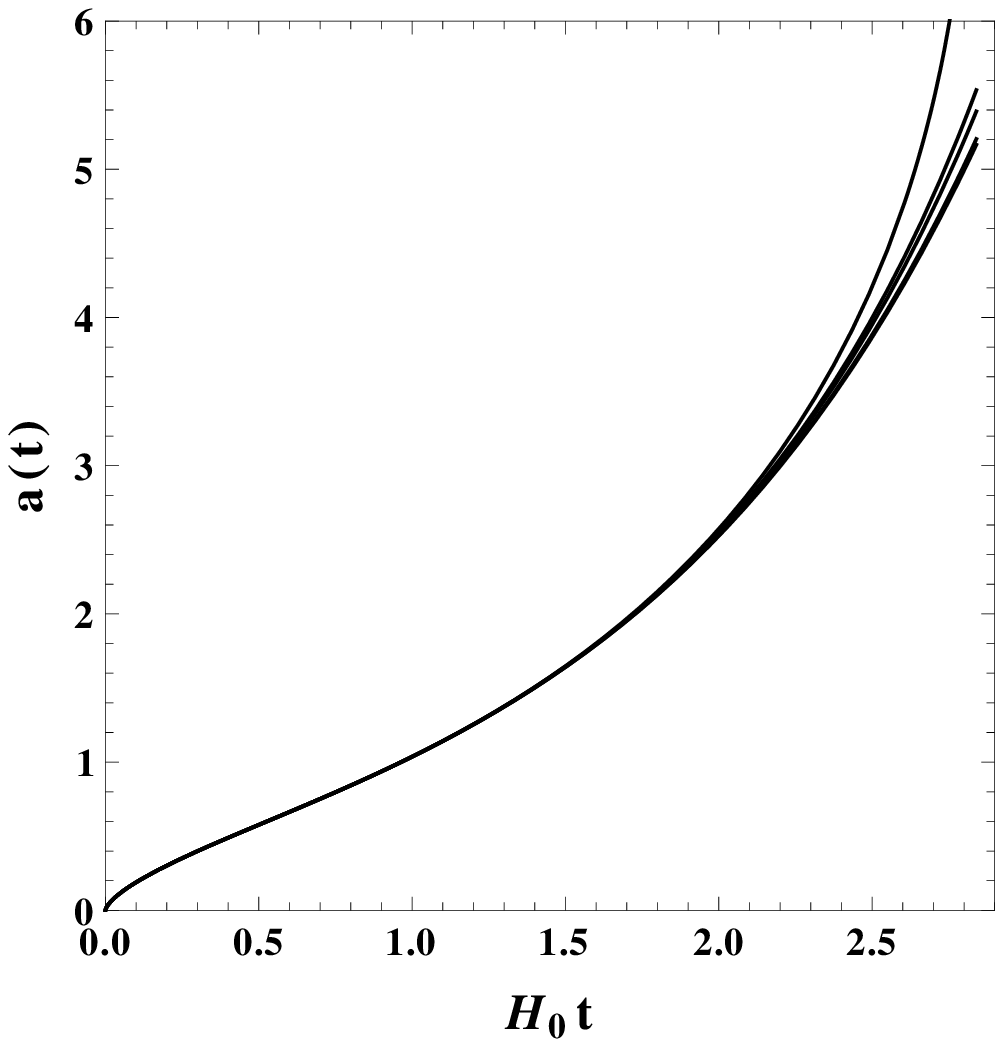}} \\ 
{\includegraphics[width=2.3in,height=2.1in,angle=0]{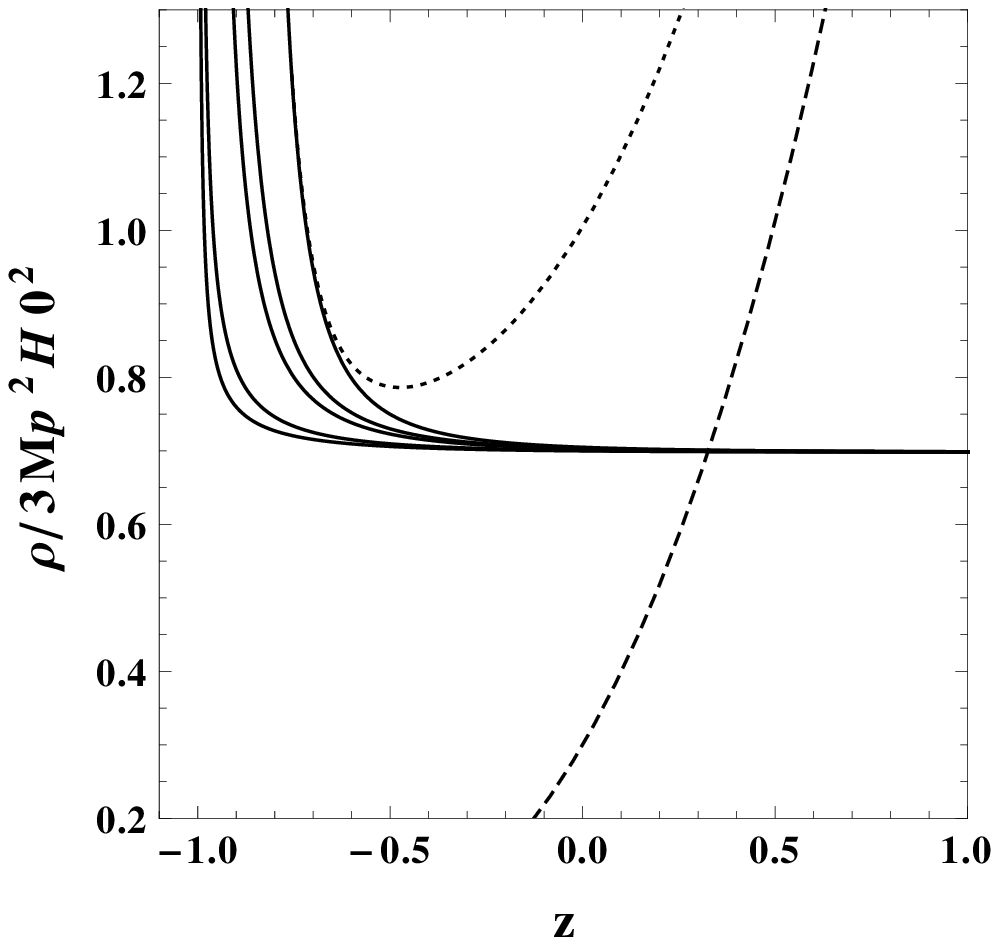}} & {%
\includegraphics[width=2.3in,height=2.1in,angle=0]{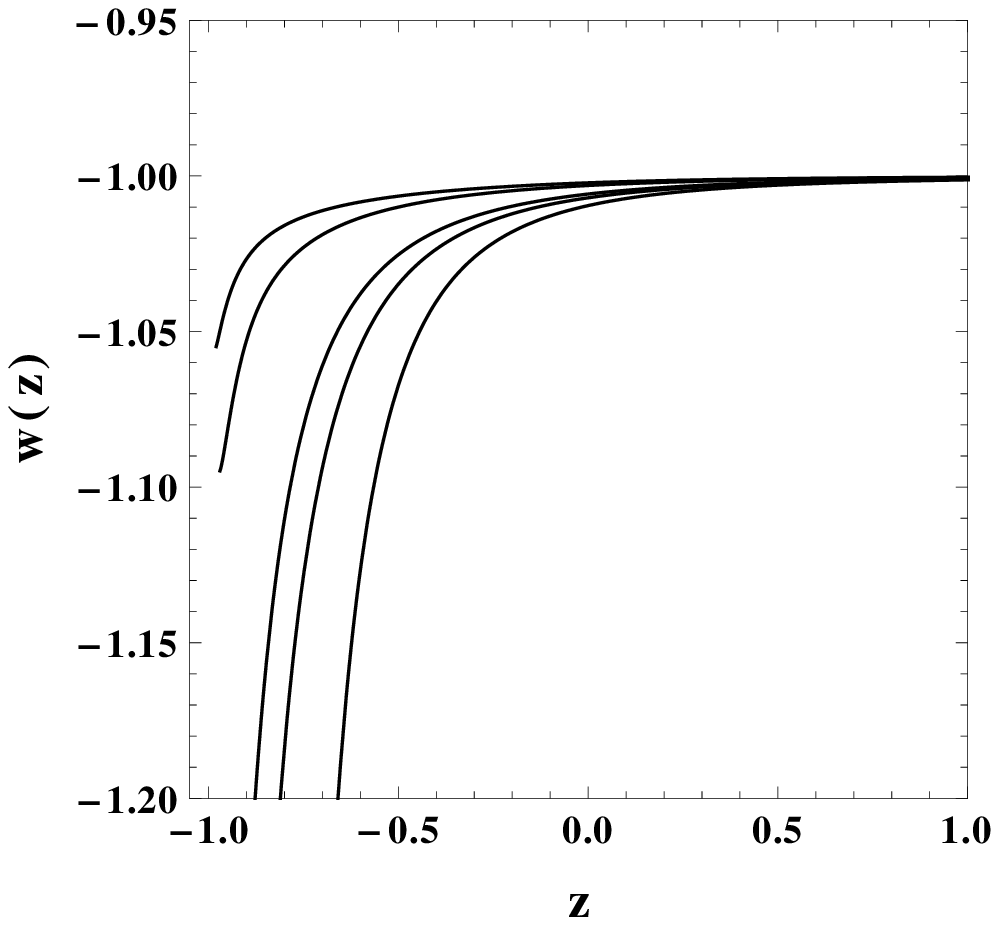}}%
\end{tabular}%
\end{center}
\caption{{\protect\small The upper panels show the evolution of potential $V(%
\protect\phi(t))$ and scale factor $a(t)$ versus time ($H_{0}t$) for
Galileon phantom field with exponential potential which is more steeper than
the linear potential for different values of $\protect\beta$. The scale
factor $a(t)$ shows the divergence nature after finite interval of time (in
distant future). Here also the time is normalized by $H_{0}$ and the present
time corresponds to $t_{0}=0.96$. The left bottom panel represents the
evolution of energy density $\protect\rho $ versus redshift $z$. The solid
lines correspond to energy density of Galileon phantom field with
exponential potential for various values of $\protect\beta$ whereas dotted
and dashed lines represent the energy density of matter and total energy
density of the Universe respectively, for standard phantom field. At late
times, the energy density of field gets to the matter, overtakes it and
begins increasing ($w<-1$), and acquires the present accelerated expansion
of the Universe. Afterwards $\protect\rho _{\protect\phi }$ continuously
blows up, in future, after a finite interval of time. The right bottom panel
shows the evolution of equation of state $w$ versus redshift $z$ for
Galileon phantom field with steep exponential potential. It has another type
of future singularity than the less steeper potential (linear potential) and
continuously blows up to $-\infty $ after definite interval of redshift. For
larger values of $\protect\beta$, the Big Rip singularity is delayed in
distant future. In this figure, all the plots have $V_0=1.2$ and $\protect%
\beta=0, 0.5, 1, 5, 10$ from top to bottom whereas bottom right plot from
bottom to top.}}
\end{figure}

\section{Galileon phantom field with exponential potential}

\label{sec:exp} We consider the Galileon phantom field with exponential
potential. This is the purely phenomenological case that is just to
establish more liberty and workability. However, this type of potential
breaks the Galileon shift symmetry. It is alike to most of the phantom field
models in which potentials are completely phenomenological. The system of
equations (\ref{eq:add1}) and (\ref{eq:phidd}) with the equation (\ref%
{eq:dimension}) for an exponential potential can be written as 
\begin{eqnarray}  \label{p7}
\frac{\ddot{a}}{a}=\dot{\phi}^{2}+\frac{3\sqrt{3}~ \beta }{2}\left( \frac{%
H_{0}}{M_{pl}}\right) ^{2}\left( \frac{\dot{a}}{a} \dot{\phi}^{3}- \dot{\phi}%
^{2}\ddot{\phi}\right)+\frac{V_{0}}{\sqrt{3}} e^{3 \phi^2} -\frac{\Omega
_{0m}}{2a^{3}}, \\
\ddot{\phi}+3 \frac{\dot{a}}{a} \dot{\phi}+ 3\sqrt{3}\beta \dot{\phi}\left( 
\frac{H_{0}}{M_{pl}}\right) ^{2}\left\{ 3\frac{\dot{a}^{2}}{a^{2}}\dot{\phi}%
+\left( \frac{a\text{ }\ddot{a}-\dot{a}^{2}}{a^{2}}\right) \dot{\phi}+2\frac{%
\dot{a}}{a}\ddot{\phi}\right\}= 2 \sqrt{3} V_0 \phi e^{3 \phi^2},
\label{p7n}
\end{eqnarray}
where, we have used $V(\phi )=V_{0}\exp ({\phi ^{2}/M_{pl}^{2}})$. Now, we
solve the system of equations (\ref{p7}) and (\ref{p7n}) numerically with
the equation (\ref{eq:init}).

For the case of $\beta=0$, the Galileon phantom field model becomes standard
phantom field model. Here we consider an exponential potential which is more
steeper than the linear one. In this potential, we obtain different type of
future singularity. The scale factor $a(t)$ diverges in distant future after
a finite interval of time as shown in the upper right panel of figure \ref%
{figexp}. In exponential potentials, Hao and Li obtained an attractor
solution having $w~ <-1$ forming the ``Big Rip'' unavoidable \cite{hao}. Our
numerical analysis shows that the energy density of phantom field blows up
after finite interval of time and correspondingly the equation of state
parameter $w$ blows up to $-\infty$ (see bottom panels of figure \ref{figexp}%
). This type of singularity has been discussed in different models, namely,
brane worlds \cite{bw}, Gauss-Bonnet cosmology \cite{gbc} and tachyonic
field \cite{tf}. For higher values of $\beta$, the Big Rip singularity is
shifted in distant future. One can say that the Big Rip singularity is
delayed for $\beta>0$.

\section{Conclusion}

In this paper, we have investigated cosmological dynamics of phantom and non
phantom fields in presence of higher derivative Galileon correction $%
\mathcal{L}_{3}$. For generality, we also studied phantom field with a
general potential in order to check the impact of Galileon term on the
structure of singularity.

In case of $\beta=0$, the Galileon field (with linear potential) reduces to
standard quintessence. In this case, as field evolves to the region of
negative values of the potential, after finite interval of time (in future),
the scale factor collapses giving rise to a Big Crunch singularity. To this
effect, energy density of Galileon field $\rho_{\phi}$ shows collapsing
nature in future. In case of standard Galileon field with linear potential,
the Big Crunch singularity can be delayed depending upon the numerical
values of $\beta$ such that for large values of the parameter, delay may be
considerable making the singularity practically redundant (see figure \ref%
{figquint}).

As for the phantom field, there are three types of singularities depending
upon the nature of potential. In case of $V\sim \phi^n$, energy density
diverges after infinite time for $n\leq 4$ whereas divergence is reached
after finite time dubbed \textit{Big Rip} if $n>4$ including the case of
exponential potential that corresponds to $n\to \infty$. In case of
potentials steeper than the standard exponential, not only divergence of
scale factor is reached in finite time but the equation of state parameter
also diverges accordingly.

We have examined the probable future regimes of Universe with Galileon
phantom field having a linear potential. In case of $\beta =0$, the Galileon
phantom field reduces to standard phantom. Due to negative sign in the
kinetic term, the field rises up along the potential giving rise to
singularity in future. The nature of this singularity is different for
different type of potentials. In case of linear potential with Galileon
phantom field, the equation of state parameter $w$ approaches $-1$ with the
slowly growing energy density compared to the standard case. For various
values of $\beta $ (0, 0.5, 1, 10, 100), we display our results in figure %
\ref{figphant} which shows that the equation of state has less and less
deviation from $\Lambda $CDM for larger values of $\beta $ and
asymptotically approaches $-1$ in distant future with the slowly increasing
energy density as shown in the bottom panels of figure \ref{figphant}. In
case of exponential potential that is more steeper than the linear one, it
has different type of singularity, the equation of state blows up to $%
-\infty $ for a definite value of redshift and correspondingly the energy
density $\rho _{\phi }$ diverges (see bottom panels of figure \ref{figexp})
that is during a definite time an infinite energy density is reached and
termed as \textquotedblleft Big Rip\textquotedblright\ singularity which
will rip galaxies apart some billion years before the actual Rip singularity
is reached \cite{0302506}. In this case, the larger values of $\beta $ will
delay the Big Rip singularity towards more and more distant future. We
therefore conclude that in general the effect of Galileon correction to
standard kinetic term in the Lagrangian generally results in the delayed
approach to singularity. It might be interesting to investigate the
behaviour of singularity using the full Galileon Lagrangian including $%
\mathcal{L}_{4}$ and $\mathcal{L}_{5}$. Again, apart from the Big crunch or
Big Rip singularities, it will be more interesting to study the effect of
Galileon correction term on the other singularities like the pressure
singularity or sudden singularity \cite{0403084} and the softer type-IV
singularity \cite{1510.04333} which were extensively been studied in \cite%
{1507.05273} and \cite{1503.08443} and is deliberated to our future
investigation.

\section*{Acknowledgements}

We thank M. Sami for his useful comments and suggestions. MS acknowledges
the financial support provided by the University Grants Commission,
Government of India, under the scheme of Dr. D. S. Kothari Postdoctoral
Fellowship. He is also thankful to M. Sajjad Athar for his constant
encouragement throughout the work. Author SKJP wish to thank Department of
Atomic Energy (DAE), Government of India for financial support through the
post-doctoral fellowship of National board of Higher Mathematics (NBHM).

\end{document}